\documentclass[10pt,a4paper,aps,amsmath,amsfonts,reprint,nofootinbib,superscriptaddress,notitlepage]{revtex4-1}

\usepackage[T1]{fontenc} \usepackage{pstricks} \usepackage{subfigure}

\usepackage{bm}
\usepackage{graphicx}
\usepackage{physics}
\usepackage{mathtools}
\usepackage{color}

\DeclareMathOperator{\sinc}{sinc}
\usepackage{physics}
\usepackage[colorlinks]{hyperref}

\hypersetup{
	bookmarksnumbered,
	pdfstartview={FitH},
	citecolor={darkgreen},
	linkcolor={darkblue},
	urlcolor={darkblue},
	pdfpagemode={UseOutlines}}
\definecolor{darkgreen}{RGB}{20,100,20}
\definecolor{darkblue}{RGB}{0,0,130}
\definecolor{darkred}{rgb}{.8,0,0}

\begin{document}

\title{Fermionic quantum carpets: From canals and ridges to solitonlike structures}

\author{Piotr T. Grochowski}
\email{piotr@cft.edu.pl}
\affiliation{Center for Theoretical Physics, Polish Academy of Sciences, Aleja Lotnik\'ow 32/46, 02-668 Warsaw, Poland}
\author{Tomasz Karpiuk}
\email{t.karpiuk@uwb.edu.pl}
\affiliation{Wydzia{\l} Fizyki, Uniwersytet w Bia{\l}ymstoku,  ul. K. Cio{\l}kowskiego 1L, 15-245 Bia{\l}ystok, Poland}
\author{Miros{\l}aw Brewczyk}
\email{m.brewczyk@uwb.edu.pl}
\affiliation{Wydzia{\l} Fizyki, Uniwersytet w Bia{\l}ymstoku,  ul. K. Cio{\l}kowskiego 1L, 15-245 Bia{\l}ystok, Poland}
\author{Kazimierz Rz\k{a}{\.z}ewski}
\email{kazik@cft.edu.pl}
\affiliation{Center for Theoretical Physics, Polish Academy of Sciences, Aleja Lotnik\'ow 32/46, 02-668 Warsaw, Poland}

\date{\today}
\begin{abstract}
We report a formation of sharp, solitonlike structures in an experimentally accessible ultracold Fermi gas, as a quantum carpet solution is analyzed in a many body system.  
The effect is perfectly coherent in a noninteracting gas, but in the presence of repulsive interaction in a two-component system, the structures vanish at a finite time.
As they disappear, the system enters a dynamical equilibrium, in which kinetic energies of atoms tend to the same average value.

\end{abstract}
\maketitle

\section{Introduction}
In 1836 Henry Fox Talbot, the father of photography, reported an unexpected result -- a diffraction grating he was observing through a magnifying lens was reappearing repeatedly in focus as he was moving away~\cite{Talbot1836}.
This phenomenon, now dubbed the \textit{Talbot effect}, was later explained by Lord Rayleigh in 1881 by means of Fresnel integrals describing near-field diffraction~\cite{LordRayleigh1881}.
It was forgotten for a long time, but nowadays its optical applications proved to be a dynamically developing branch of physics, involving numerous realizations~\cite{Patorski1989,Wen2013}.

The Talbot effect is a consequence of an interference of highly coherent waves and it is not surprising that there exists its quantum counterpart.
Quantum revivals~\cite{Eberly1980}, quantum fractals~\cite{Berry1996,Wojcik2000}, quantum echoes~\cite{Buchkremer2000}, quantum Talbot effect~\cite{Sanz2007}, and quantum scars~\cite{Heller1984,Kaplan1998} are all closely connected manifestations of the time evolution of wave packets~\cite{Robinett2004}.

\begin{figure*}[hbt]
	\includegraphics[width=0.47\linewidth]{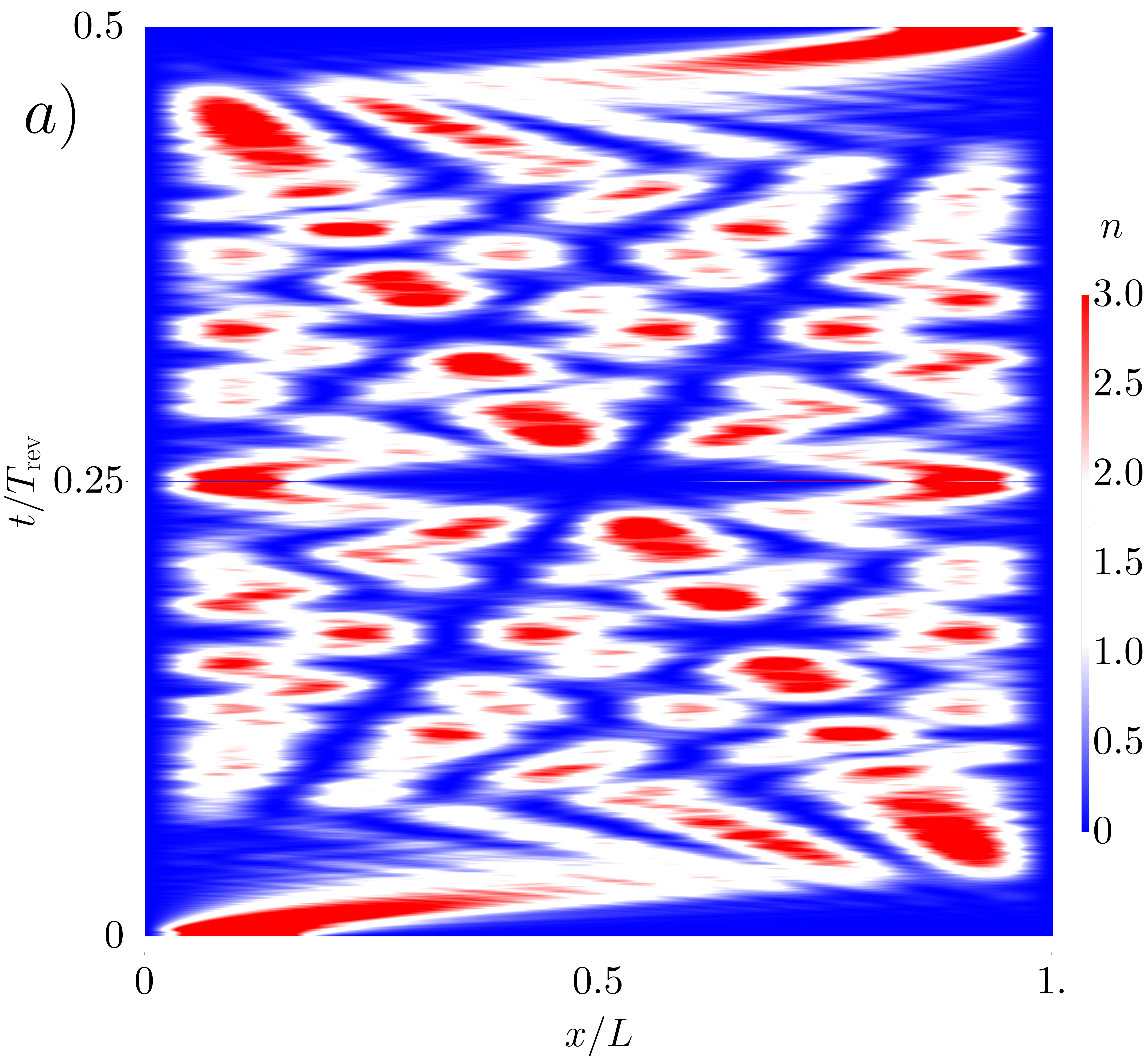}
	\includegraphics[width=0.49\linewidth]{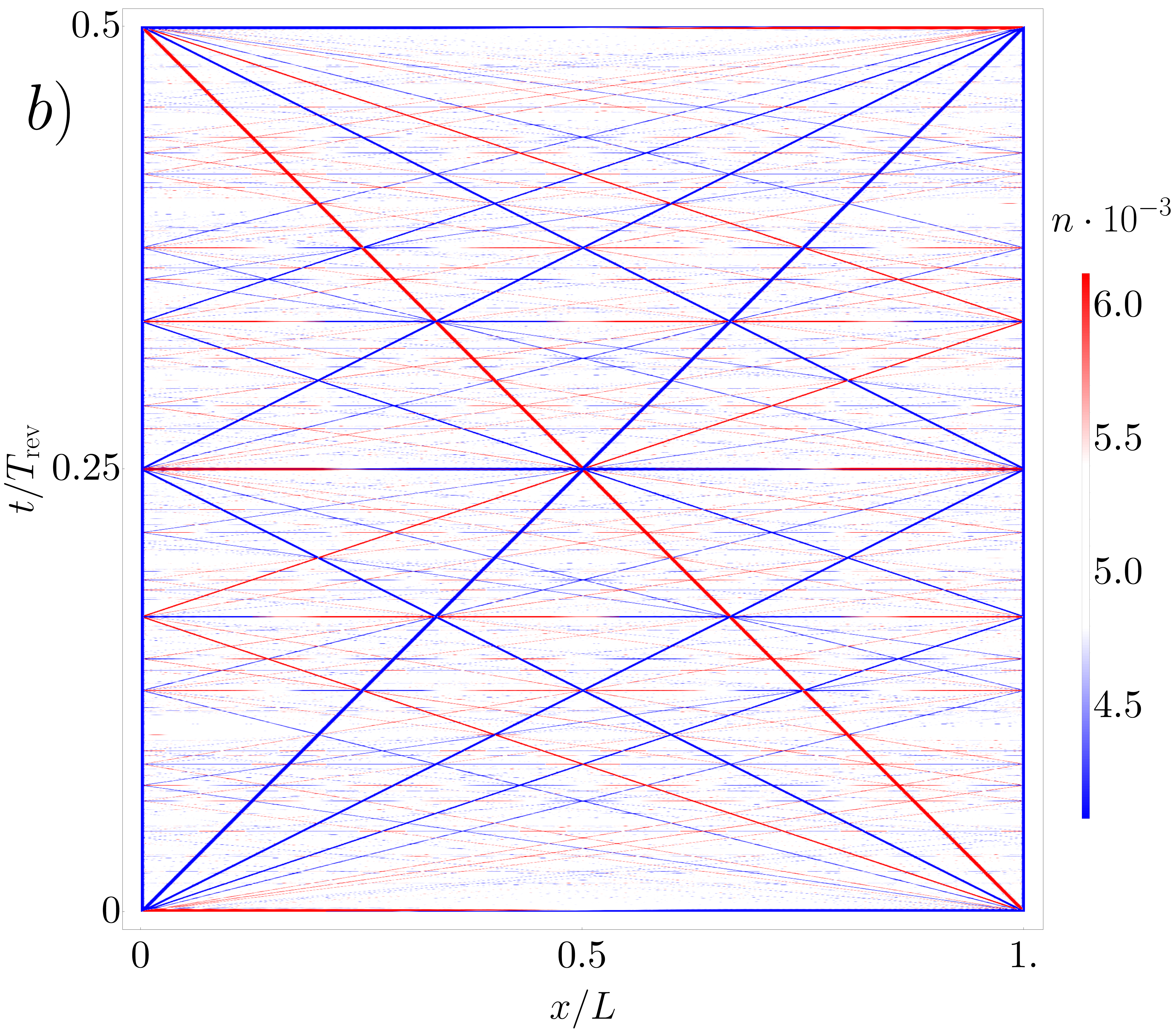}
	\caption{ Radially-integrated probability density plots for one atom (a) and 5000 atoms (b) initially confined to a box with the width of $D/L=0.21$ and a perpendicular harmonic trap.
		In contrast to usual one atom situation in which blurry canals and ridges emerge, scenario involving larger number of particles is characterized by much sharper features.
		These structures become solitonlike -- thin, localized, and shape-preserving.
		\label{comp1}}
\end{figure*}

In this Letter, we focus on aesthetically appealing \textit{quantum carpets} -- spatiotemporal representations of a probability density of a quantum particle in a box.
Firstly observed by Kinzel~\cite{Kinzel1995}, then named and heavily studied by Schleich and coworkers~\cite{Stifter1997,Grossmann1997,Marzoli1998,Kaplan1998a,Berry2001}, they stand out due to characteristic structures, called \textit{canals} and \textit{ridges} (see Fig.~\ref{comp1}(a)).
These patterns do not follow classical trajectories and origin only from interference terms.
They have been studied from various perspectives, including Wigner representation~\cite{Marzoli1998,Friesch2000}, degeneracy in intermode traces~\cite{Grossmann1997,Stifter1997,Marzoli1998,Friesch2000,Loinaz1999}, travelling wave decomposition~\cite{Hall1999}, spin chains~\cite{Banchi2015,Genest2016a,Genest2016,Lemay2016,Kay2016,Compagno2016,Christandl2017}, and fractional revivals~\cite{Aronstein1997}, emphasizing their deep links to number theory~\cite{Berry2001}, quantum computing~\cite{Harter2001,Harter2001a}, decoherence effects~\cite{Kazemi2013}, and factorization of numbers through Gauss sums~\cite{Berry1996a,Wolk2011,Bigourd2008}.

\begin{figure}[hbt]
		\includegraphics[width=0.95\linewidth]{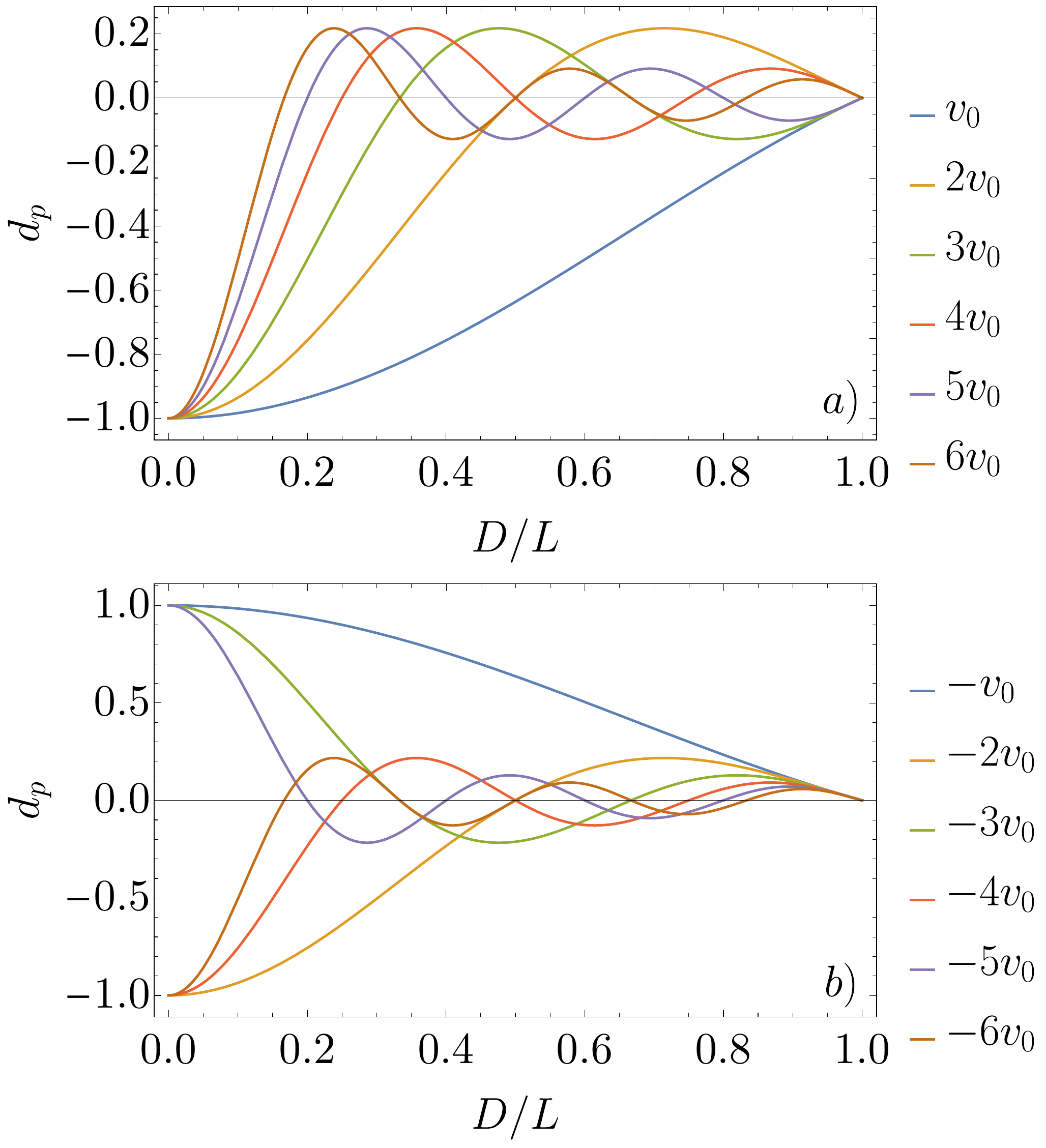}
				\caption{ Relative depths of solitonlike structures, both for right- and left-movers.		\label{comp3}}
	\end{figure}

Up to date, the experimental realizations of self-reviving systems by means of spatial Talbot interferometry are numerous - they span from atoms~\cite{Nowak1997,Chapman1995,Cronin2009} and molecules~\cite{Brezger2002,Hornberger2012}, through electrons~\cite{Sonnentag2007,Korneev2012} and light~\cite{Case2009} to Bose-Einstein condensates~\cite{Ryu2006}.
Its temporal counterpart, closer to the quantum carpet situation, has also been investigated -- examples include ultracold bosonic gases~\cite{Deng1999,Mark2011}, Rydberg states~\cite{Ahn2001}, and nuclear wave packets~\cite{Katsuki2009}.

However, quantum carpets were discussed almost exclusively in bosonic systems - whether it was light or a Bose-Einstein condensate~\cite{Ruostekoski2001,Gawryluk2006}.
The interest in many-body fermionic systems was scarce~\cite{Nest2006}, mostly due to the difficulty of considering highly correlated particles.
Nonetheless, we show that even in the limit of ideal gas of polarized fermions in an infinite well some interesting phenomena arise.

We show that degenerate Fermi gas that is initially trapped in a box and then released into a bigger one exhibits solitonlike structures, that move analogously to canals and ridges from the one-particle problem (see Fig.~\ref{comp1}(b)).
These structures are characterized by a constant relative depth in density as a number of atoms grows, effectively making them more pronounced in larger systems.
This feature is however absent when different initial trapping potentials (e.g. harmonic one) are considered.
Moreover, we show that this phenomenon is not destroyed by temperature and should be available in a gas of distinguishable particles.
Despite the fact that our starting point is a one-dimensional gas, three-dimensional scenario is also explored, revealing access to experimentally achievable regimes.

As a next step of our considerations, we investigate two-component repulsive Fermi gas that interact via s-wave collisions~\cite{Jo2009,Sommer2011,Valtolina2016,Trappe2016,Grochowski2017a,Stoner1933,Massignan2014,Cui2014,Jiang2016}.
If both components are initially trapped in different parts of the box and then released, solitonlike structures are present in both species separately.
However, due to the repulsive interspecies interaction, they start to diminish in time -- the faster, the stronger the repulsion is -- until they ultimately vanish.

\section{Ideal Fermi gas}
We start our considerations with an ideal polarized Fermi gas.
We assume that at the beginning of the evolution, the many-body wave function of $N$ indistinguishable fermionic atoms is given by the single Slater determinant:
$\Psi (x_1,...,x_{N}) = \frac{1}{\sqrt{N!}} \det \left( \phi_1, \cdots,\phi_N \right)$, where $\phi_i(x),\, {i=1,...,N}$ denote different, orthonormal orbitals.
The gas is then released to evolve freely in a box trap with the length of $L$.
Eigenfunctions of such a box potential are standing waves:
$\varphi_k(x) = \sqrt{2/L} \sin{\left( k \pi x/L  \right) } \theta{\left( x \right)} \theta{ \left( L- x \right) }, k=1,2,\ldots,$
where $\theta(x)$ is a Heaviside step function and eigenenergies read $E_k=k^2 \pi^2 \hbar^2/2 m L^2$.
Let us introduce overlaps between initial orbitals and box trap eigenfunctions, $\lambda(n,k) \equiv (\varphi_k,\phi_n)$.
We consider noninteracting gas, so there is no mixing between different orbitals as they undergo a unitary evolution, $\phi_n(x,t)=\sum_{k=1}^{\infty} \lambda(n,k) \varphi_k (x) \exp\left( {-i E_k t / \hbar}\right) $.
We can write down the time evolution of the orbitals squared, separating contributions moving with different velocities:
\begin{align} \label{exall}
&\left| \phi_n(x,t)\right|^2  \approx   \sum_{k=1}^{\infty} \lambda^2(n,k) \varphi_k^2 \ - \nonumber \\
&\sum_{p} \sum_{k=1}^{\infty} \frac{1}{L} \lambda(n,k) \lambda(n,k+|p|)\cos{\left( (2k+|p|)\frac{\pi}{L} (x- p v_0 t)\right) },
\end{align}  
where $v_0 = \pi \hbar/2 m L$ is the characteristic velocity of the box that is connected to time of the system's revival $T_{\text{rev}}=2 L/ v_0 = 4 L^2 m/\pi \hbar$.
As we can see, we can fully describe such a system in terms of travelling contributions that move with the velocities that are multiples of $v_0$. $p\in \mathbb{Z} \setminus \{0\}$ will denote each of these terms.

The first term in Eq.~\ref{exall} is independent of time and constitutes a background for a time evolution.
Therefore, a formula for $p$-th contribution in the system of $N$ fermions reads:
\begin{align}
n_p(x,t)  \approx & - \sum_{n=1}^{N} \sum_{k=1}^{\infty} \frac{1}{L} \lambda(n,k) \lambda(n,k+|p|) \times \nonumber \\
&\cos{\left( (2k+|p|)\frac{\pi}{L} (x- p v_0 t)\right) }.
\end{align}

We note that for each orbital, $p$-th contribution is peaked at $x_0=p v_0 t$ for a right-moving one ($p>0$) and at $x_0=L+p v_0 t$ for a left-moving one ($p<0$).
Such a behavior manifests itself as canals and ridges in the one particle problem and can be explained by the interference terms in the Wigner representation.
Therefore, we evaluate these contributions at appropriate peaks and introduce relative depth of each one:
\begin{align}
d_p =& \frac{n_p (x_0,t)}{n}  = \sigma(p) \frac{1}{N} \sum_{n=1}^{N} \sum_{k=1}^{\infty}  \lambda(n,k) \lambda(n,k+|p|), \nonumber  \\ 
\sigma(p) = & 
\begin{cases} 
-1, \ \  & p>0, \\
(-1)^{mod(|p|,2)+1},\ \  & p<0
\end{cases}
\end{align}
where $n=N/L$ is the average density of the fermionic cloud.

Following the calculation featured in the Appendix, we can write the whole $p$-th relative depth as
\begin{align}\label{dens32}
d_p = -\sigma(p) \sqrt{\frac{2}{\pi}} \frac{1}{N} \mathcal{F}^c\{n(x,0)\} \left(\frac{|p| \pi}{L}\right),
\end{align}
where $\mathcal{F}^c\{n(x,0)\}$ is a cosine Fourier transform of the initial one-particle density of the fermionic gas.
To evaluate it in the limit of a large $N$, we can use LDA.
By $v(x)$ we will denote a potential that initially traps fermions and therefore implies initial orbitals $\phi_n$.
The one-particle density of such a gas in Thomas-Fermi approximation reads $n(x)=\sqrt{2 m}/\pi \hbar \sqrt{\mu-v(x)}\theta(\mu-v(x))$, where $\mu$ is a chemical potential.
Therefore, the relative depth $d_p$ has an approximate dependence on $N$, $d_p \sim \sqrt{\mu (N)}/N$.
In here we assume that initially the whole cloud lies within the box trap with the length of $L$ - it is approximately satisfied whenever classical turning points of potential $v(x)$ for a particle of Fermi energy lie within the box.
The $N$-dependence can be obtained through the normalization condition, $\label{norma} N=\int_{0}^{L} \dd{\tilde{x}} \tilde{n}(\tilde{x}) \rightarrow \mu(N)$.
It is helpful to note the connection of this dependence to the energy levels of the initial fermionic cloud.
For a general potential $v(x)$ we can approximate generated energy levels by the WKB quantization condition, 
$n + C =\sqrt{2 m}/\pi \hbar \int_{-\infty}^{\infty} \dd{x}  \sqrt{E_n-v(x)}\theta(E_n-v(x))$, where $C$ is a constant depending on the boundary conditions in turning points.
This equation has the exactly same form as the normalization condition, so it yields the same results -- the dependence of the WKB energy spectrum on the quantum number $n$ is identical to $N$-dependence of the chemical potential of the fermionic cloud.

We look for situations in which at least one of the moving contributions $n_p$ has a constant shape in space and preserves its depth with a growing number of atoms in the system.
One of the necessary conditions to satisfy the latter is for $d_p \sim \sqrt{\mu (N)}/N$ to be constant in the limit $N \rightarrow \infty$.
However, it does not mean that every such a initial wavepacket would behave as needed -- it only guarantees that contribution from every orbital is of the same magnitude.

It also shows why initial harmonic confinement would not generate distinct solitonlike contributions -- their depth would scale like $1/\sqrt{N}$, making them disappear for large number of atoms.
We will focus on the simplest system that at the beginning has a quadratic spectrum -- ideal Fermi gas confined to a box trap.
At first, atoms are trapped in a infinite well with the length of $D$, that is smaller than the box to which they are released, $D<L$.
Both traps share one of the walls.

Firstly, we find the $p$-th relative depth for each orbital, $d_p^n$, by performing explicitly Fourier transform:
\begin{align}
d_p^n = \sigma(p) \sinc{\left(D |p| \pi / L\right)} \frac{4 n^2 \pi^2}{ \left( 2 n \pi \right)^2 -\left( D|p| \pi /L \right)^2 }.
\end{align}
As we can see, for large $N$, each of these contributions is the same, meaning that absolute depth of the moving terms grows linearly with $N$.
The relative depth is therefore
\begin{align}
d_p =\sigma(p) \sinc{\left(D |p| \pi / L\right)}.
\end{align}
It is also interesting to note that we get the same result by inserting appropriate Thomas-Fermi profile into~\eqref{dens32}.
Moreover, we also reproduce it by numerical evaluation of the exact expression~\eqref{exall}, using exact overlaps between considered modes.

Firstly, we will make sure that our candidate for a stable time evolution indeed preserves its shape during the evolution.
In Fig.~\ref{comp1}(a, b) we compare density plots for one fermion and 5000 of them, but with additional perpendicular trapping.
For one atom case, canals and ridges are clearly visible.
In the second case, they are visible as well, however they have become much sharper -- thinner and more pronounced.
Each of the moving contributions is now a sharply peaked solitonlike structure that preserves its shape during evolution and is characterized by a constant velocity.
In Fig.~\ref{comp3} we plot relative depths of such structures, both for right- and left-moving contributions.

\section{Three-dimensional setup}
To explore experimental accessibility of fermionic quantum carpets, we now proceed to consider three-dimensional geometry.
The trapping in $x$-axis remains unchanged as compared to 1D scenario, but in perpendicular directions we assume arbitrary confinement.
In the Appendix we find the approximate formula for a $p$-th contribution to the one-particle density integrated over perpendicular degrees of freedom in $T=0$:
\begin{align}
n_p \approx n d_p \sinc{\left( \eta k_F (x- p v_0 t) \right) },
\end{align}
where $k_F$ is a Fermi wavevector of the gas in the initial confinement and $\eta$ is a parameter that is found numerically for each type of perpendicular trapping.
This approximation works well close to $x_0$, as far as $|x-x_0| \sim \pi / \eta k_F$ and it gives a very good estimation of the width of the structure, that is the same for each of contributions:
\begin{align}
w=w_p=w_0 / \eta k_F,
\end{align}
where $w_0 \approx 3.79$.
In case of 1D system, the approximation is almost exact, with $\eta=2$ and $k_F=\pi N / D$.
Such a scaling means that the structures become extremely thin and operationally unreachable by standard imaging in the quasi-1D systems.
However, full 3D system is much more promising -- e.g. for the perpendicular trapping with the oscillator length of $a_\perp$, Fermi wavevector reads $\left( 15 \pi N D^{-1} a_\perp^{-4} \right) ^{1/5}$, $\eta \approx 1.3$ and the structures are characterized by the width of several microns in experimentally accessible systems.

Moreover, we check what happens to considered solitonlike structures in the presence of nonzero temperature by taking the Fermi-Dirac distribution into account.
It is easy to analytically show that their depth is unaffected and we numerically confirm that they become thinner by up to 50\% in $T=4 T_F$.
Existence of these structures even in high temperatures suggests that considered effect should be visible even for a gas that is governed by classical, Boltzmann distribution, but in the quantum system with a quantized spectrum.

\section{Two-component repulsive Fermi gas}
We now proceed to consider a repulsive two-component Fermi gas.
We stick to the simple description of the single determinant Hartree Fock Ansatz for the wave function.
Such a description misses quantum corrections due to interspecies correlations, but it captures qualitatively one-body behavior in a weakly interacting regime that we consider (see e.g.~\cite{Grochowski2017a}).
As the gas has now two components, two according spin states are introduced.
$\phi_n (x)$ now denote not orbitals, but orthonormal spin-orbitals and $x$ comprise of both spatial and spin degrees of freedom.
We assume that spin-dependent part of $\phi_n (x)$ is twofold and the same numbers of atoms occupy each spin state.
\begin{figure}[htbp!]
	\centering
	\includegraphics[width=0.49\linewidth]{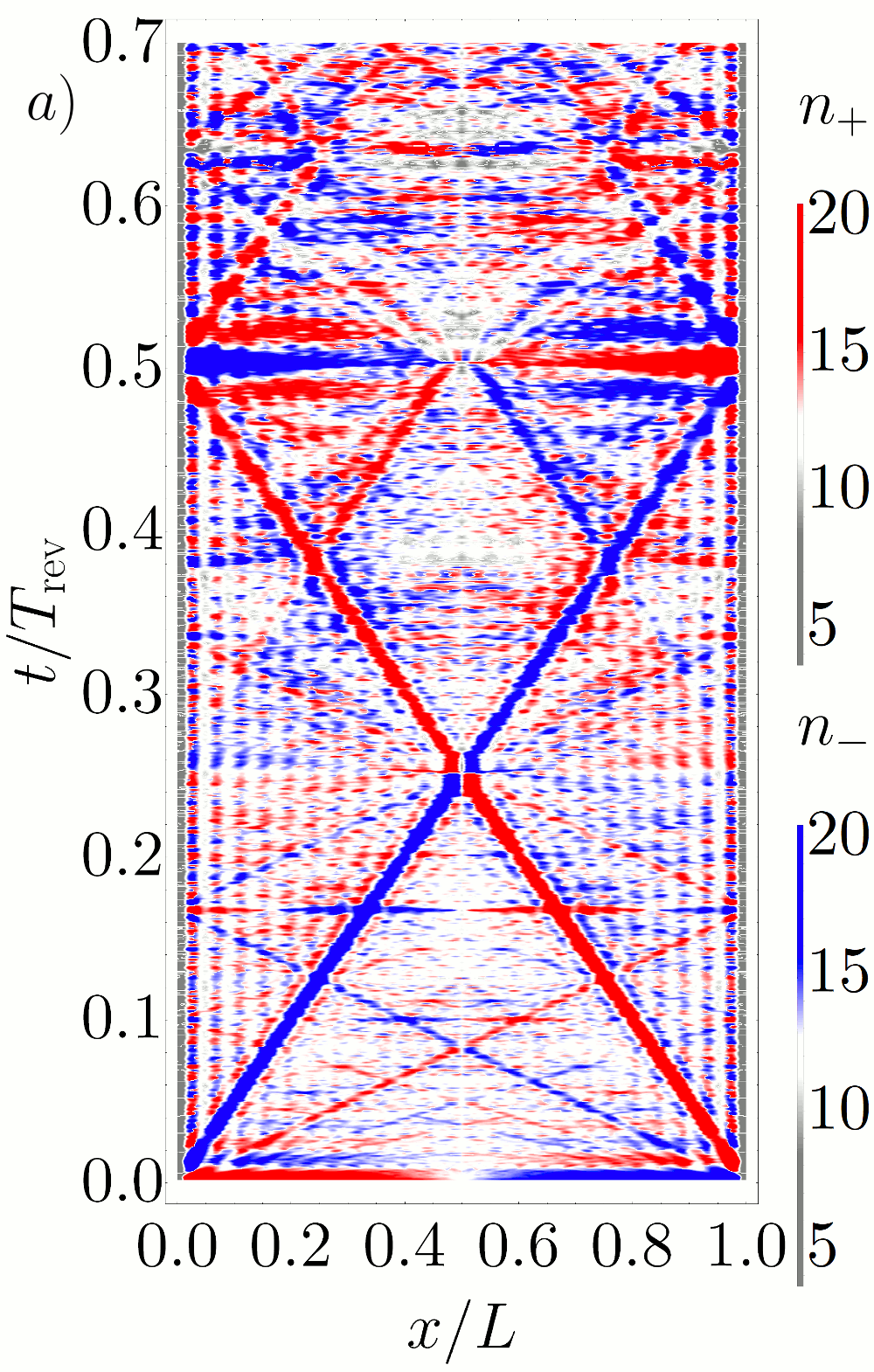}
	\includegraphics[width=0.49\linewidth]{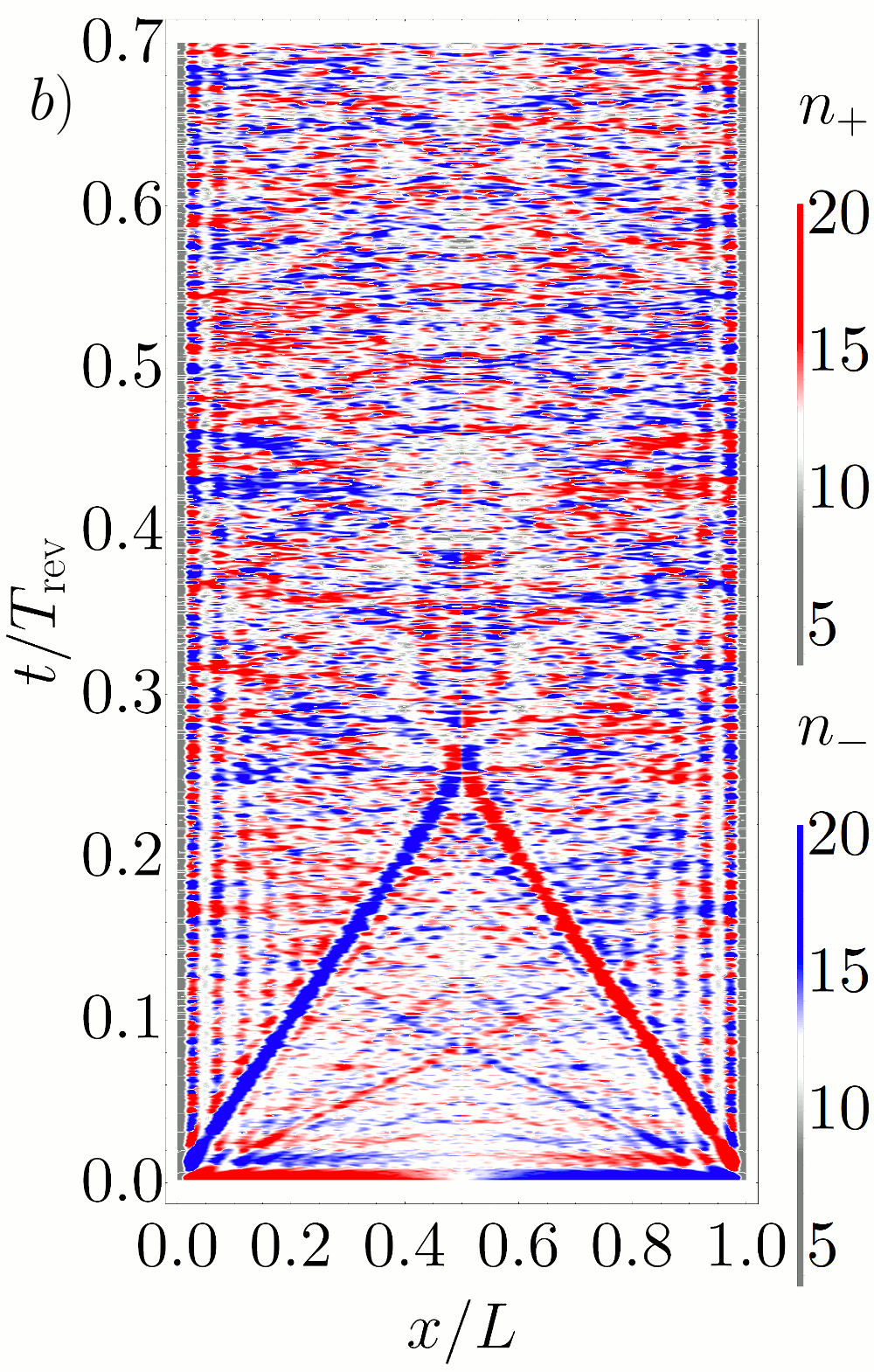}
	\caption{Comparison between the spatiotemporal density evolution of $12+12$ atoms initiated in a separated state for different values of contact interaction: $g=16$ (a) and $g=32$ (b) in one dimension.
		The solitonlike structures are characterized by a finite lifetime in the presence of the interaction.
		\label{comp5}}
\end{figure}

\begin{figure}[htbp!]
	\centering
	\includegraphics[width=1.0\linewidth]{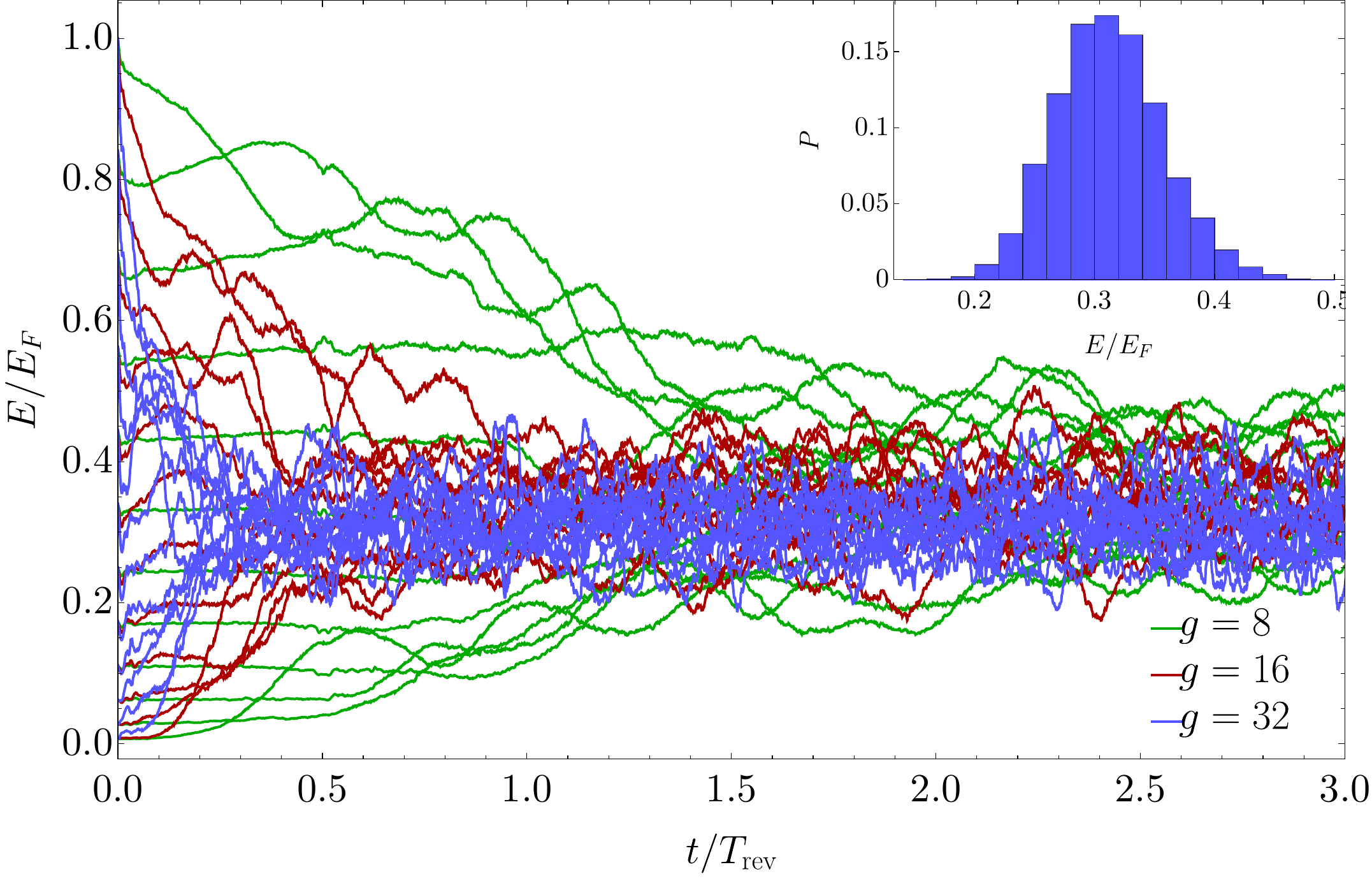}
	\caption{ Kinetic energies for consecutive spin-orbitals of one of the species as a function of time for different values of interaction strength.
		After some critical time, the dynamical equilibrium is reached -- there is no distinguished ordering of the orbitals as their kinetic energies fluctuate around a common average value.
		Time-averaged distribution of the kinetic energies proves to be positively skew (see the histogram for $g=32$ in the inset).
		The time needed to achieve equilibrium is of the same order as the lifetime of the solitonlike structures visible in the one-particle density of the gas.
		\label{comp6}}
\end{figure}

\begin{figure*}[htbp!]
	\centering
	\includegraphics[width=0.8\linewidth]{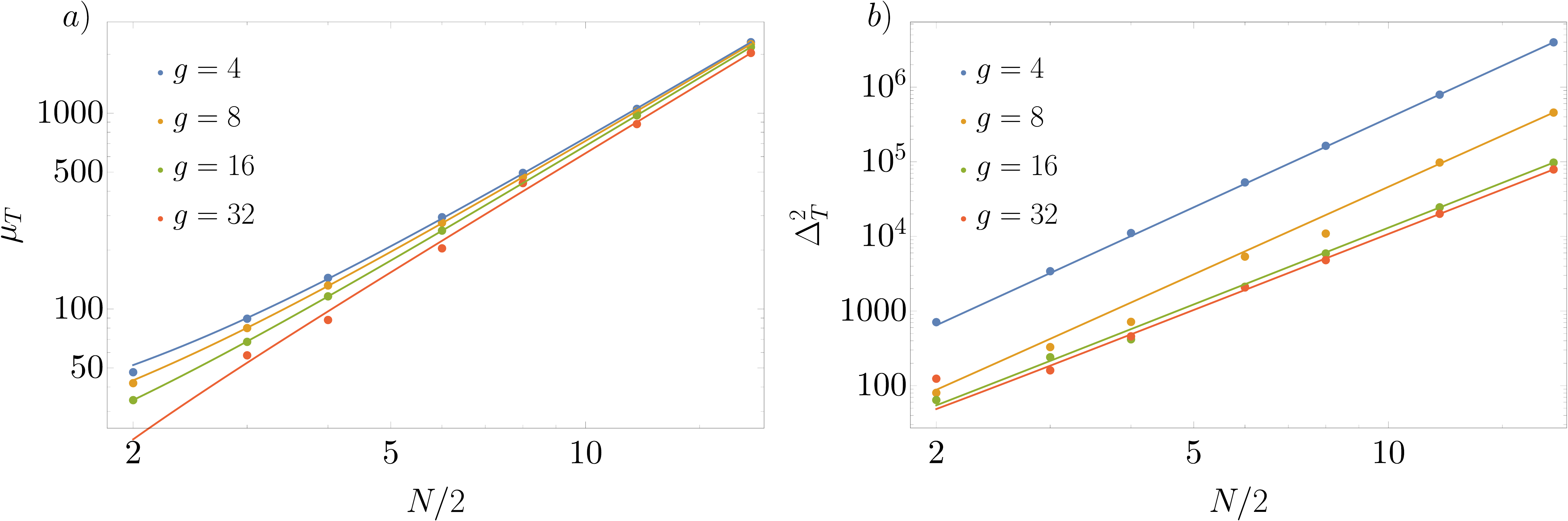}
	\caption{ The power-law scaling for the mean kinetic energy $\mu_T$ (a) and the kinetic energy variance $\Delta^2_T$ (b) as a function of the number of atoms in the system.
		Fitted power-law curves are included.
		\label{wn2}}
\end{figure*}

In this description, atoms in each spin state can be considered a noninteracting Fermi gas with the only interaction present being an interspecies one.
We model this interaction by a repulsive contact potential, characterized by a nonnegative coupling constant $g \geq 0$.
The dynamics of such a system is provided by the means of time-dependent Hartree-Fock equations~\cite{Gawryluk2017}. (see more in the Appendix)

We analyze situation in which two species are initially separated by a thin barrier in the middle of a box potential that traps both of them.
Then, gas is released from within these initial walls to evolve freely in a larger box.
For no interaction, two fermionic quantum carpets unfold symmetrically with an infinite lifetime and infinite full revivals.
As the interaction is turned on, the lifetime of coherent evolution becomes finite and the solitonlike structures eventually disappear (see Fig.~\ref{comp5}).
The fade-out of the structures occurs when the system enters a dynamical equilibrium, as the kinetic energies of the spin-orbitals become roughly equal during the evolution in time (see Fig.~\ref{comp6}).

Equilibration in such a system has not been thoroughly studied. 
However, systems that are Bethe Ansatz-solvable (and this system is one of those \cite{Oelkers2006}) usually equilibrate (but not necessarily thermalize) under so-called weak eigenstate thermalization hypothesis (ETH) \cite{Jensen1985,Deutsch1991,Srednicki1994,DAlessio2016,Deutsch2018,Mori2018}.
This feature was rigorously shown for the translationally invariant systems \cite{Biroli2010,Iyoda2017}, but as in our system such an invariance is not present, it is not immediately clear whether this is the case.
However, the equilibrated state is observed and we can characterize its finite-size scaling, in the spirit of the usual ETH analysis~\cite{Beugeling2014}.

\subsection{Analysis of the statistical properties of the equilibrated state}
We perform analysis of the distribution of the kinetic energies of the spin-orbitals after gas enters the equilibrated state.
In Fig.~\ref{wn2} we plot dependence of the mean kinetic energy $\mu_T$ and the variance $\Delta^2_T$ on the number of atoms in the system.
The analysis is performed for different values of interaction strength.
The scalings read $\mu_T \sim N^2$ for the mean and $\Delta^2_T \sim N^{3.4-3.9}$ for the variance.
Such a scaling of the mean is well retrieved for all checked values of the interaction.
In the case of the variances, the power-law behavior is retrieved, however quality of the fit is not as good due to finite length of time evolution analyzed.
For smaller values of the interaction, the scaling law tends to be higher (up to $\sim N^{3.9}$) and for stronger interaction it is closer to $\sim N^{3.4}$.

Dependence of the mean kinetic energy $\mu_T$ and the variance $\Delta^2_T$ on the interaction strength can be analogously studied.
Again, the power-law fit is better for the mean than it is for the variance.
However, scaling laws can be relatively unambiguously determined as $\mu_T \sim -g$ and $\Delta^2_T \sim g^{-0.5}$.

\subsection{Analysis of the coherence in the system}
\begin{figure*}[htbp!]
	\centering
	\includegraphics[width=1.0\linewidth]{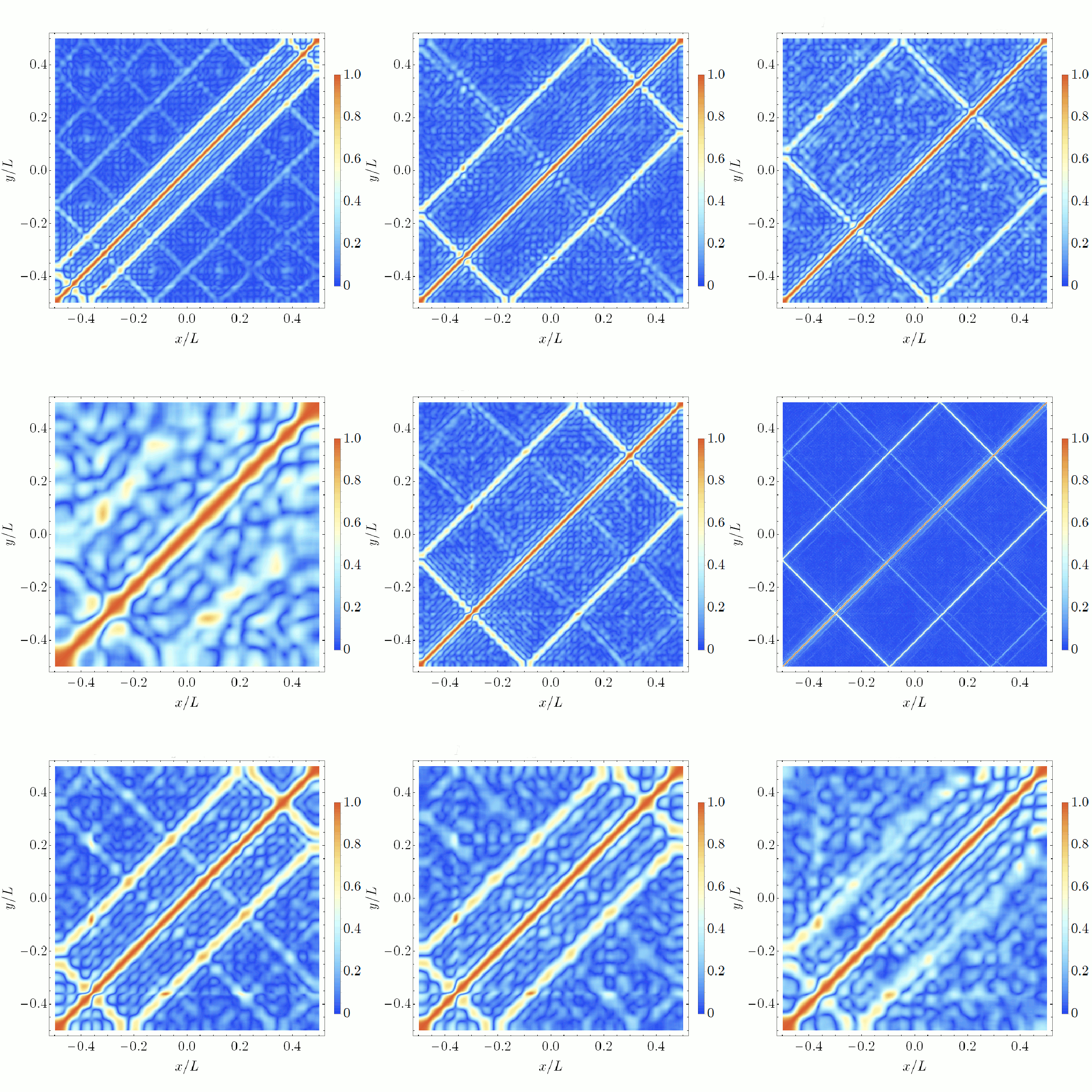}
	\caption{ The modules of first oder correlation function, $|g^{(1)}|$ for a fermionic quantum carpet of two-component gas presented in the main text.
		In each of the figures rectangle-shaped structures in off-diagonal terms are clearly visible.
		Each of the single-particle density solitonlike structure has its counterpart in such a representation.
		It is the most distinguishably seen in the top left figure, in which several such rectangle-shaped structures are present.
		The visibility of each of the structures is proportional to the relative depth of associated solitonlike structure.
		We present their behavior with a changing time, number of particles in the system and interaction.
		In the top row, from left to right, consecutive instances of time evolution are shown ($t=\left\lbrace 0.031,0.086,0.14\right\rbrace  T_{\text{rev}}$ with $N=24+24$, $g=24$ kept constant).
		Taking the most pronounced structure as a reference point, one can see that each structure evolves in time, changing shape, but keeping the slope and overall perimeter constant, giving the impression of 'breathing'.
		In ideal gas, such an evolution is infinite as the coherence is not lost.
		However, in a finite interaction situation, the structures get dimmer through time evolution.
		In the central row, different numbers of atoms are presented ($N=\left\lbrace 8+8, 24+24, 96+96 \right\rbrace$ with $g=24$, $t=0.1 T_{\text{rev}}$ kept constant).
		With growing $N$, the structures get thinner and the background becomes less fluctuating.
		Finally, in the bottom row, effect of interaction is shown -- the stronger the interaction is, the faster the structures vanish ($g=\left\lbrace 2, 16, 32 \right\rbrace$ with $N=12+12$, $t=0.071 T_{\text{rev}}$ kept constant).
		\label{comp_app1}}
\end{figure*}

It is worth noting that even for very small number of atoms (3+3), each scalings can be retrieved.
Moreover, it is important to underline the fact that these results are not intrinsic for the analyzed Hamiltonian, as the evolution in Hilbert space is restricted by the choice of single-Slater Ansatz for the wavefunction.

\begin{figure*}[htbp!]
	\centering
	\includegraphics[width=1.0\linewidth]{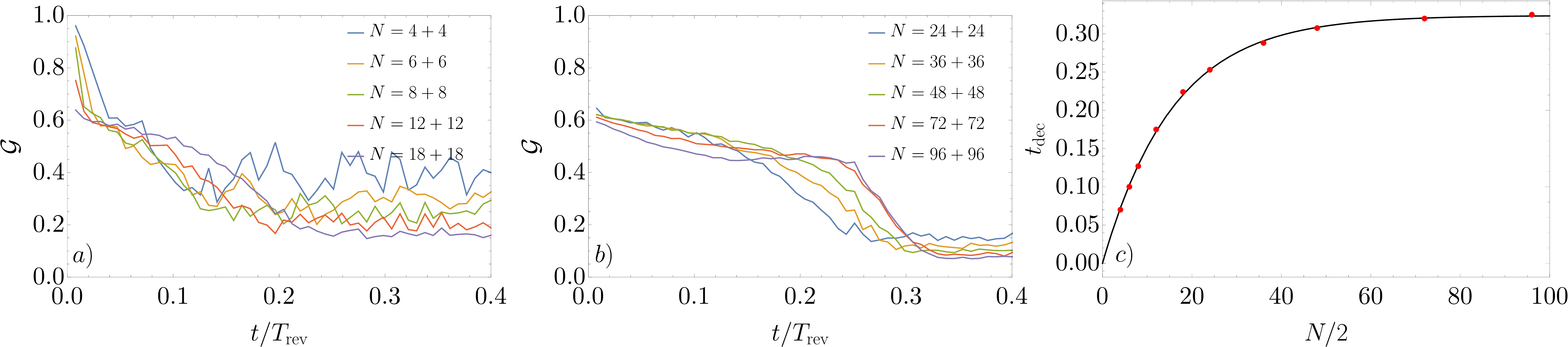}
	\caption{(a) and (b) The coherence measure $\mathcal{G}$ as a function of time for different number of atoms in the system for a constant coupling constant $g=24$.
		(c) Time of decoherence $t_{\text{dec}}$ extracted from two other figures with a fitted curve.
		\label{times}}
\end{figure*}

\begin{figure*}[htbp!]
	\centering
	\includegraphics[width=1.0\linewidth]{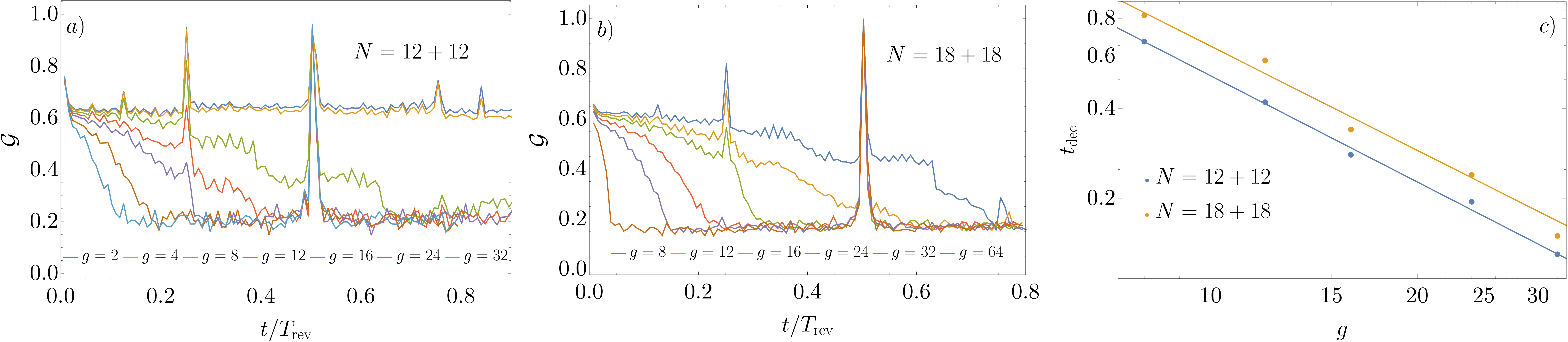}
	\caption{ (a) and (b) The coherence measure $\mathcal{G}$ as a function of time for different values of interaction with a constant number of particles -- $N=12+12$ and $N=18+18$.
		The visible spikes in coherence occur every half the revival time due to reducing of integration path of $\mathcal{G}$ to the diagonal.
		(c) Time of decoherence $t_{\text{dec}}$ extracted from two other figures with fitted curves.
		\label{gs}}
\end{figure*}

%\begin{figure*}[htbp!]
%	\centering
%	\includegraphics[width=0.8\linewidth]{wg2.pdf}
%	\caption{ The power-law scaling for the mean kinetic energy $\mu_T$ (left) and the kinetic energy variance $\Delta^2_T$ (right) as a function of the interaction.
%		Fitted power-law curves are included.
%		\label{wg2}}
%\end{figure*}

%Under the equilibrium, kinetic energies of the orbitals fluctuate around common value, which we find out to scale like $\sim N^2$ and $\sim -g$.
%Furthermore, their variances scale like $\sim N^{3.4-3.9}$ and $g^{-0.5}$. (Calculations can be found in Supp. Mat.)

Initially, existence of macroscopic structures is due to strong coherence present in the system.
Their fade-out during the evolution can be explained and quantitatively described by the progressing loss of the coherence.
To analyze this loss, we propose a measure of coherence that aims to quantify their dimming during the time evolution.
The starting point of our considerations is normalized first order correlation function for one of the species:
\begin{align}
g^{(1)}(x,y,t)=g^{(1)}_+(x,y,t)=\frac{\left\langle \widehat{\Psi}_+^\dagger (x,t) \widehat{\Psi}_+ (y,t) \right\rangle }{\sqrt{n_+(x,t)} \sqrt{n_+(y,t)}}.
\end{align}
We only consider one of the species, as the initial condition implies time evolution of the second component to be symmetric with respect to the center of the trap.
The function $g^{(1)}$ is normalized such that its value is 1 on the diagonal.
In Fig.~\ref{comp_app1} we present the results of calculations of this first order of coherence as a function of different parameters.
The caption below this Figure fully addresses the explanation of the results and the introduction of the rectangular-shaped structures in the $g^{(1)}$ .

It can be shown that the height of each of the rectangular-shaped structures is roughly equal to the relative depth $d_p$ of the associated solitonlike stucture.
However, due to intrinsic fractal behavior of these structures and the following relatively large fluctuations, we have chosen to use modulus of $g^{(1)}$ function averaged over the whole first rectangular-shaped structure as a measure of coherence:
\begin{align}
\mathcal{G}\left( t \right)  \equiv  \int_{-L/2}^{x_0(t)} \dd x \left| g^{(1)} \left( x, -x-L/2+x_0(t) \right)  \right| + \nonumber \\ \int_{x_0(t)}^{L/2} \dd x \left| g^{(1)} \left( x, x-L/2-x_0(t) \right)  \right|,
\end{align}
where $x_0(t)$ is an vertex of the rectangle at $y=-L/2$, and is equal to:
\begin{align}
x_0(t) =  \begin{cases} 
L/2 - 2 t/T_{\text{rev}} + \lfloor 2 t/T_{\text{rev}} \rfloor & \text{if} \ \lfloor 2 t/T_{\text{rev}} \rfloor \  \text{is odd} \\
-L/2 + 2 t/T_{\text{rev}} - \lfloor 2 t/T_{\text{rev}} \rfloor & \text{if} \ \lfloor 2 t/T_{\text{rev}} \rfloor \  \text{is even}
\end{cases}
\end{align}

The structures are characterized by a constant velocity of 'breathing' in the noninteracting case, and this velocity is almost not affected in the interacting one.
However, small fluctuations of this velocity happen.
It is completely negligible in smaller systems for which the width of the structures is relatively big, nonetheless for larger numbers of atoms, as the structures get thinner, some more precise treatment in order to find a maximum of coherence peak is needed.
To deal with this problem, we evaluate correlation function at couple of points near the expected trajectory and include the biggest one into the integral.
This way, we can effectively follow the maxima of coherence for a given time.

It is worth noting that in the noninteracting case, $\mathcal{G}$ is almost a constant function of time that equals $d_1$.
In a case analyzed here, we use initial width of the cloud to be $D/L=0.5$, so $\mathcal{G}_0=d_1=2/\pi$.
We use this measure to evaluate time of decoherence for which the structures effectively die off.
In Fig.~\ref{times} the results of the analysis are presented in the case in which the number of atoms in the system is increased.
For every case, after the system equilibrates, the value of $\mathcal{G}$ saturates at some value.
It is likely due to the finite size of the system -- the saturated value always decreases with the size of the system.
For small number of atoms, the curve strongly fluctuates, and as this number gets larger, the fluctuations are getting smaller, making the curve almost smooth.
We made calculations up to $N=96+96$ atoms in the system (keeping interaction $g$ constant) and it seems that time of decoherence initially grows with the number of atoms, and then saturates at some value.
The curve that fits best to these results has a form of $t_{\text{dec}} \sim 1-\exp(-c N)$, with $c$ being some positive constant.

In Fig.~\ref{gs} the dependence on the interaction is presented.
While keeping number of atoms constant, time dependence of $\mathcal{G}$ is plotted for different values of the coupling constant $g$.
The stronger the interaction is, the faster the decoherence happens -- it scales like $t_{\text{dec}} \sim g^{-1.2}$.

\section{Recapitulation}

%We propose a measure of coherence (defined and further elaborated in Supp. Mat.) that allows us to determine timescales in which solitonlike structures are lost.
%We find that this time of decoherence is diminished with the interaction strength, scaling like $~\sim g^{-1.2}$ and not according to power law in number of particles.
%For $N$ up to $96+96$ the best fit is an exponential, $\sim 1-\exp(-c N)$.
%Again, such a behavior persists even for very small number of atoms, revealing universality of such a scaling.

To recapitulate, we have found previously unobserved phenomenon, closely connected to the quantum carpet spatiotemporal profile.
In a large system consisting of ultracold fermions, very sharp, solitonlike structures appear for which we find analytical description of their velocities, depths, and widths.
We stress out that the connection to the solitons known from nonlinear equations is only due to its apparent behavior -- e.g. the constant shape and velocity. 
Underlying mechanism of creation of these structures is fundamentally different and does not guarantee the robustness under small perturbations, unlike in the classically known soliton.
Furthermore, we analyze the effect of repulsive interaction between two spin species of Fermi gas on the presence of this phenomenon.
We study the decoherence of the solitonlike structures due to interactions and their disappearance for a stronger repulsion.
As a future line of work, it is worth to further analyze equilibration in such a system, with a potentially interesting additions of random fields or different geometries.
Another potential route would involve studying systems that map onto noninteracting fermions, e.g. strongly interacting bosons.

\begin{acknowledgments}
P.T.G. was supported by (Polish) National Science Center Grant 2018/31/N/ST2/01429.
T.K. and M.B. were supported by (Polish) National Science Center Grant 2018/29/B/ST2/01308.
K.R. was supported by (Polish) National Science Center Grant 2015/19/B/ST2/02820.
\end{acknowledgments}

\appendix
\onecolumngrid
\appendix
\section{Derivation of $p$-th contribution}
\setcounter{equation}{0}
Let's focus on the $p$-th contribution coming from each orbital:
\begin{align}
d_p^n =  \sigma(p) \sum_{k=1}^{\infty}  \lambda(n,k) \lambda(n,k+|p|).
\end{align}
Firstly, we recall the Fourier transform
\begin{align}
\hat{f}(k) = \frac{1}{\sqrt{2 \pi}} \int_{-\infty}^{\infty} f(x) e^{i k x} \dd{x},
\end{align}
and assuming that we will consider only real functions, we introduce sine and cosine transforms:
\begin{align}
\hat{f}^s(k) = \mathcal{I} \hat{f}^(k), \ \ \ \hat{f}^c(k) = \mathcal{R} \hat{f}^(k).
\end{align}
We can immediately see that $\hat{f}^s(k) = - \hat{f}^s(-k)$.
Let's evaluate overlaps between initial orbitals and box eigenmodes:
\begin{align}
\lambda(n,k) = \int_{-\infty}^{\infty} \dd{x} \ \varphi_k^\star (x) \phi_k (x) = \sqrt{\frac{2}{L}} \int_{-\infty}^{\infty} \dd{x} \phi_n(x) \sin{\left( \frac{k \pi}{L} x\right)}=\sqrt{\frac{4 \pi}{L}}\widehat{\phi_n}^s\left(\frac{k \pi}{L}\right),
\end{align}
where $\phi_n(x)=\phi_n(x) \theta{\left( x \right)} \theta{ \left( L- x \right) }$ is meant to be truncated into the box with the width of $L$.
Therefore, we can write
\begin{align}
d_p^n = & \sigma(p) \sum_{k=1}^{\infty}  \frac{4 \pi}{L} \widehat{\phi_n}^s\left(\frac{k \pi}{L}\right) \widehat{\phi_n}^s\left(\frac{(k+|p|) \pi}{L}\right) \longrightarrow \nonumber \\
& 2  \sigma(p) \int_{-\infty}^{\infty} \dd{\tilde{k}} \widehat{\phi_n}^s\left(\tilde{k}\right) \widehat{\phi_n}^s\left(\frac{|p| \pi}{L}-\tilde{k}\right) =  2  \sigma(p) \  \widehat{\phi_n}^s \star \widehat{\phi_n} ^s\left(\frac{|p| \pi}{L}\right),
\end{align}
where $\star$ denotes usual convolution that in our case takes form:
\begin{align}
\hat{f}^s \star \hat{g}^s(l)&=\frac{1}{2 \pi}\int_{-\infty}^{\infty} \dd{k} \int_{-\infty}^{\infty} \dd{x} \int_{-\infty}^{\infty} \dd{y} f(x)\sin(kx) g(y) \sin((l-k)y) \nonumber \\
&=-\frac{1}{2}  \int_{-\infty}^{\infty} \dd{x} f(x) g(x) \cos(lx) + \frac{1}{2}  \int_{-\infty}^{\infty} \dd{x} f(x) g(-x) \cos(lx)
\end{align}
However, we consider only functions that vanish outside $x \in [0,L]$, so the above expression can be simplified into
\begin{align}
\hat{f}^s \star \hat{g}^s(l) =-\frac{\pi}{2} \widehat{f g }^c (l).
\end{align}
We arrive at the compact form for the $p$-th relative depth for a given orbital:
\begin{align}\label{reldeporb}
d_p^n =   2  \sigma(p) \  \widehat{\phi_n}^s \star \widehat{\phi_n} ^s\left(\frac{|p| \pi}{L}\right) = -\sigma(p) \sqrt{\frac{2}{\pi}} \widehat{\phi_n^2}^c \left(\frac{|p| \pi}{L}\right).
\end{align}
We can therefore write the whole $p$-th relative depth as
\begin{align}
d_p = -\sigma(p) \frac{1}{N} \sum_{n=1}^{N} \sqrt{\frac{2}{\pi}} \widehat{\phi_n^2}^c \left(\frac{|p| \pi}{L}\right) = -\sigma(p) \sqrt{\frac{2}{\pi}} \frac{1}{N} \widehat{\sum_{n=1}^{N} \phi_n^2}^c \left(\frac{|p| \pi}{L}\right) = -\sigma(p) \sqrt{\frac{2}{\pi}} \frac{1}{N} \widehat{n(x,0)}^c \left(\frac{|p| \pi}{L}\right),
\end{align}
where $n(x,0)$ is the initial one-particle density of the fermionic gas.

\section{Derivation of widths of the structures}
\setcounter{equation}{0}
In three dimensions, atoms are trapped in a box in $x$-direction and in an arbitrary perpendicular confinement:
\begin{align}
V(x,y,z)=\text{Box}(x)+V_y(y)+V_z(z).
\end{align}
As such, orbitals are now characterized by three independent quantum numbers, $n=(n_x,n_y,n_z)$.
Analogously to 1D case, we consider single-particle density, but integrated over perpendicular degrees of freedom:
\begin{align}
\left| \phi_n(x,t)\right|^2 = \int \dd z \int \dd y \left| \phi_n(x,y,z,t)\right|^2 = \left| \sum_{k=1}^{\infty} \lambda(n_x,k) \varphi_k (x) e^{-i E_k t / \hbar} \right|^2.
\label{ert}
\end{align}
Expression \eqref{ert} differs from its one-dimensional counterpart by changing $n$ to $n_x$ inside a sum.
Again, we introduce $p$-th contribution:
\begin{align}
n_p(x,t)  \approx - \sum_{n=1}^{N} \sum_{k=1}^{\infty} \frac{1}{L} \lambda(n_x,k) \lambda(n_x,k+|p|)\cos{\left( (2k+|p|)\frac{\pi}{L} (x- p v_0 t)\right) },
\label{sua}
\end{align}
but this time we can use explicit expressions for overlaps, as we have been considering a box potential in $x$-axis:
\begin{align}
\lambda(n,k)=
\begin{cases} 
\frac{2}{\pi} \sqrt{\frac{D}{L}}  (-1)^{n+1}\sin\left(\frac{k \pi D}{L}\right)  \frac{n}{n^2-\left( \frac{kD}{L}\right)^2 }, \ \  & n L \neq k D, \\
\sqrt{\frac{D}{L}}  ,\ \  & n L = k D.
\end{cases}
\end{align}
One can explicitly check that multiplication of functions $\lambda$ in \eqref{sua} can be approximated by
\begin{align}
\lambda(n_x,k) \lambda(n_x,k+|p|) \approx \frac{D}{L} \operatorname{sinc}\left( \frac{k \pi D}{L} - n_x\pi \right) \operatorname{sinc}\left( \frac{(k+|p|) \pi D}{L} - n_x\pi \right),
\end{align}
and is centered around
\begin{align}
k_0 = \frac{n_x L}{D} - \frac{|p|}{2}.
\end{align}
As a next step, we stick to region close to the structure's peaks, working with variables describing distance from them, $x_p=(x- p v_0 t)-x_0^p$.
Then, we identify slowly varying parts of $p$-th contribution in this region:
\begin{align}
n_p(x,t)  \approx & - \sum_{n=1}^{N} \sum_{k=1}^{\infty} \frac{1}{L} \frac{D}{L} \operatorname{sinc}\left( \frac{k \pi D}{L} - n_x\pi \right) \operatorname{sinc}\left( \frac{(k+|p|) \pi D}{L} - n_x\pi \right) \cos{\left( (2k+|p|)\frac{\pi}{L} (x- p v_0 t)\right) } \nonumber \\
\approx &-\sum_{n=1}^{N} \frac{1}{L} \cos{\left( (2k_0+|p|)\frac{\pi}{L} x_p\right) } \sum_{k=1}^{\infty} \frac{D}{L} \operatorname{sinc}\left( \frac{k \pi D}{L} - n_x\pi \right) \operatorname{sinc}\left( \frac{(k+|p|) \pi D}{L} - n_x\pi \right).
\label{asd}
\end{align}
Sum over $k$ in \eqref{asd} can be turned into an integral, that can be readily estimated:
\begin{align}
\int \dd k \ \frac{D}{L} \operatorname{sinc}\left( \frac{k \pi D}{L} - n_x\pi \right) \operatorname{sinc}\left( \frac{(k+|p|) \pi D}{L} - n_x\pi \right) = \operatorname{sinc}\left( \frac{\pi D}{L} |p| \right).
\end{align}
Within these approximations, we are left with expression for $p$-th contribution to the single-particle density:
\begin{align}
n_p(x_p,t) \approx &-\sum_{n=1}^{N} \frac{1}{L} \cos{\left( (2k_0+|p|)\frac{\pi}{L} x_p\right) } \operatorname{sinc}\left( \frac{\pi D}{L} |p| \right) = -\frac{1}{L} \operatorname{sinc}\left( \frac{\pi D}{L} |p| \right) \sum_{n=1}^{N} \cos{\left(2 n_x \pi \frac{  x_p}{D}\right) }.
\end{align}
An expression
\begin{align}
\sum_{n=1}^{N} \cos{\left(2 n_x  \pi \frac{ x_p}{D}\right) }
\label{hard}
\end{align}
can be readily calculated in one dimension, where 
\begin{align}
n_x(n)=n, \ \ \ \ n_{\text{max}}=N.
\end{align}
In this case, it is called Langrange formula and yields
\begin{align}
\sum_{n=1}^{N} \cos{\left(2 n  \pi \frac{ x_p}{D}\right) } = -\frac{1}{2}+\frac{\sin{\left( \left(N+\frac{1}{2} \right) 2 \pi \frac{x_p}{D}  \right) }}{2 \sin{\left(\frac{\pi x_p}{D} \right) }} \approx - N \sigma(p) \operatorname{sinc}\left(2 k_F x_p \right),
\end{align}
where $k_F$ is the initial Fermi wavevector of the gas:
\begin{align}
k_F = \frac{\pi N}{D}
\end{align}
Therefore, $p$-th contribution can be invoked in form
\begin{align}
n_p(x_p) \approx \frac{N}{L} \sigma(p)  \operatorname{sinc}\left( \frac{\pi D}{L} |p| \right)  \operatorname{sinc}\left(2 k_F x_p \right) = \frac{N}{L} d_p  \operatorname{sinc}\left(2 k_F x_p \right).
\end{align}
However, a sum \eqref{hard} cannot be explicitly calculated in three-dimensional case, but we find out that $p$-th contribution can be numerically approximated by
\begin{align}
n_p(x_p)\approx \frac{N}{L} d_p  \operatorname{sinc}\left(\eta k_F x_p \right),
\end{align}
where $\eta$ is a constant that depends on the character of perpendicular trapping.
The approximation is reasonably accurate for $x_p$ up to
\begin{align}
|x_p| \sim \frac{\pi}{\eta k_F}.
\end{align}
With this approximation of the shape, we can calculate widths of the half maximum for each solitonlike structure:
\begin{align}
w = \frac{w_0}{\eta k_F}, \ \ \ w_0=3.79098...
\end{align}
For different types of trappings, we have differents values of $k_F$ and $\eta$.
For box trap with the length of $D_y$ in $y$-direction and box trap with the length of $D_z$ in $z$-direction:
\begin{align}
k_F=\left(\frac{3}{4} \pi^2 \frac{N}{D D_y D_z} \right)^{1/3}, \ \ \ \ \eta \sim 3.2.
\end{align}
For harmonic trap with the length of $a_y=\sqrt{\frac{\hbar}{m \omega_y}}$ in $y$-direction and box trap with the length of $D_z$ in $z$-direction:
\begin{align}
k_F=\left(16 \pi \frac{m \omega_y}{\hbar}\frac{N}{D D_z} \right)^{1/4}, \ \ \ \ \eta \sim 1.4.
\end{align}
For harmonic trap with the length of $a_y=\sqrt{\frac{\hbar}{m \omega_y}}$ in $y$-direction and harmonic trap with the length of $a_z=\sqrt{\frac{\hbar}{m \omega_z}}$ in $z$-direction:
\begin{align}
k_F=\left(15 \pi \frac{m \omega_y}{\hbar} \frac{m \omega_z}{\hbar} \frac{N}{D} \right)^{1/5}, \ \ \ \ \eta \sim 1.3
\end{align}

\section{Numerical method}
\label{num_method}

In order to study a repulsive two-component Fermi gas in a one-dimensional space we approximate the many-body wave function of the system of $N$ indistinguishable fermionic atoms by the single Slater determinant

\begin{eqnarray}
&&\Psi ({\bf x}_1,...,{\bf x}_{N})
= \frac{1}{\sqrt{N!}} \left |
\begin{array}{lllll}
\phi_1({\bf x}_1) & . & . & . & \phi_1({\bf x}_{N}) \\
\phantom{aa}. &  &  &  & \phantom{aa}. \\
\phantom{aa}. &  &  &  & \phantom{aa}. \\
\phantom{aa}. &  &  &  & \phantom{aa}. \\
\phi_{N}({\bf x}_1) & . & . & . & \phi_{N}({\bf x}_{N})
\end{array}
\right |  .   \nonumber  \\
\label{Slater}
\end{eqnarray}

The coordinates ${\bf x}_n$ ($n=1,...,N$) of atoms include both spatial and spin variables and $\phi_n({\bf x})$ $(n=1,...,N)$ mean the orthonormal spin orbitals. Since we consider a two-component Fermi gas, the spin-dependent part of the spin-orbitals is twofold. We further assume that equal number of atoms occupy each spin state.

Atoms occupying the same spin state are considered as a noninteracting Fermi gas. The only interaction present in the system is the one between different spin atoms. It is described by the contact potential with the coupling constant equal to $g$. For such spin-dependent interactions, the time-dependent Hartree-Fock equations for the spatial parts of the spin-orbitas, $\phi_n^{+}(x,t)$ and $\phi_n^{-}(x,t)$, can be written as

\begin{eqnarray}
&&i\hbar \frac{\partial}{\partial t}  \phi_n^{+} (x,t) =
\left( -\frac{\hbar^2}{2 m} \frac{\partial^2}{\partial x^2} + V_{tr}({x})  
+ g\, n^{-} (x,t) \; \right) \; \phi_n^{+} (x,t)
\nonumber  \\
&&i\hbar \frac{\partial}{\partial t}  \phi_n^{-} (x,t) =
\left( -\frac{\hbar^2}{2 m} \frac{\partial^2}{\partial x^2} + V_{tr}({x})  
+ g\, n^{+} (x,t) \; \right) \; \phi_n^{-} (x,t)
\label{HFeq}
\end{eqnarray}
for $n=1,...,N/2$. The atomic densities of components (normalized to the number of atoms), $n^{+} (x,t)$ and $n^{-} (x,t)$, are defined as follows 
\begin{eqnarray}
&&   n^{+} (x,t) = \sum_{j=1}^{N/2} |\phi_j^{+} (x,t)|^2         \nonumber  \\
&&   n^{-} (x,t) = \sum_{j=1}^{N/2} |\phi_j^{-} (x,t)|^2  \,.
\end{eqnarray}

Initially, $N$ atoms occupy the lowest energy states of one-dimensional boxes, positioned side by side, each of length $L/2$. At $t=0$ the internal wall is removed and the dynamics starts. We use the split-operator method to solve Eqs. (\ref{HFeq}) [54]. The spatial step of the numerical grid is $\Delta x=0.0025 L$ and we choose the time step as $\Delta t=10^{-7}\, mL^2/\hbar$. Eqs. (\ref{HFeq}) have two constants of motion, both the total number of atoms and the total energy of the system $(E)$ are preserved. We monitor these quantities while solving the evolution equations (\ref{HFeq}). They are both conserved to very high accuracy, namely $\Delta N /N < 10^{-8}$ and $\Delta E /E < 10^{-5}$ at the end of calculations.

\twocolumngrid

\end{document}